\documentclass{article}
\usepackage{spconf,amsmath,graphicx}
\usepackage{cite}
\usepackage{amssymb,mathtools}
\usepackage{booktabs,adjustbox,siunitx,xcolor,colortbl,etoolbox}
\usepackage[symbol]{footmisc}
\usepackage{enumerate,enumitem}

\usepackage[absolute]{textpos}

\TPMargin{5pt}

\newcommand{\copyrightstatement}{
    \begin{textblock}{0.84}(0.08,0.93)    
         \noindent
         \footnotesize
         \copyright 2021 IEEE.  Personal use of this material is permitted.  Permission from IEEE must be obtained for all other uses, in any current or future media, including reprinting/republishing this material for advertising or promotional purposes, creating new collective works, for resale or redistribution to servers or lists, or reuse of any copyrighted component of this work in other works.
    \end{textblock}
}

\title{NOISY ORACLE PAPER TITLE}
\title{Noise-aware Training Criteria for Speech Separation with Noisy Oracles}
\title{Training Models Single-Channel Speech Separation Models with Noisy Supervision}
\title{Training Noisy Single-Channel Speech Separation with Noisy Oracle Sources: A Large Gap and a Small Step}
\name{Matthew Maciejewski$^{1,2}$, Jing Shi$^{1,3}$, Shinji Watanabe$^{1,2}$, Sanjeev Khudanpur$^{1,2}$}
\address{$^1$ Center for Language and Speech Processing, The Johns Hopkins University, USA \\
         $^2$ Human Language Technology Center of Excellence, The Johns Hopkins University, USA \\
         $^3$ Institute of Automation, Chinese Academy of Sciences, China
}
\begin{document}
\copyrightstatement
\ninept
\maketitle
\begin{abstract}
As the performance of single-channel speech separation systems has improved, there has been a desire to move to more challenging conditions than the clean, near-field speech that initial systems were developed on. When training deep learning separation models, a need for ground truth leads to training on synthetic mixtures. As such, training in noisy conditions requires either using noise synthetically added to clean speech, preventing the use of in-domain data for a noisy-condition task, or training using mixtures of noisy speech, requiring the network to additionally separate the noise. We demonstrate the relative inseparability of noise and that this noisy speech paradigm leads to significant degradation of system performance. We also propose an SI-SDR--inspired training objective that tries to exploit the inseparability of noise to implicitly partition the signal and discount noise separation errors, enabling the training of better separation systems with noisy oracle sources.
\end{abstract}
\begin{keywords}
speech separation, noisy speech, deep learning
\end{keywords}
\section{Introduction}
\label{sec:intro}

In recordings of speech with multiple people present, such as the case of conversational speech, it is common for the speech signals to overlap, as people talk simultaneously~\cite{bengio2005machine,mtgoverlap}. Most speech technologies are designed to work on only a single speaker's speech, suffering a degradation of performance in the overlapping speech condition~\cite{msmeetings}. It can even be difficult for human listeners to understand this speech. Speech separation aims to solve this problem by producing multiple waveforms from a single mixture, each containing speech from only one speaker.

In the advent of the proliferation of deep learning, the performance of speech separation has been improved greatly~\cite{tasnet,dprnn,wavesplit,shi2020sep}, leading to a desire to build better systems that are robust to a wide variety of conditions~\cite{whamr,mx6ch5,librimix} beyond the clean, near-field data~\cite{dpcl_wsj02mix} that systems were initially developed on. Due to deep learning models requiring ground truth signals, systems are trained using synthetic mixtures; and, accordingly, noisy speech separation systems have typically been trained and evaluated on mixtures where the noise has been added synthetically as well, to great success~\cite{wham,whamr,wavesplit,deepcasa}. However, the use of synthetic noise prevents using any speech data with existing noise. Recordings of speech considered to be clean rarely exist outside of recording studios, with many practical applications including some level of noise~\cite{spmag}, effectively disallowing in-domain training using this paradigm.

In cases where access to clean speech is not possible, such as the CHiME challenges~\cite{chime5,chime6}, there has been little success in using single-channel speech separation systems. In addition, in cases with synthetic mixtures of real noisy data, performance has been shown to suffer\cite{mx6ch5}. It is unfortunately impossible to make perfect comparisons across these data paradigms---the standard evaluation metrics~\cite{sisdr,sdr,stoi,pesq} require ground truth and are affected in the same manner as the systems are in training. However, the evidence available suggests there may truly be a gap in performance.

While many real conditions include reverberation---an additional challenge---in this paper, we focus our investigation on the noisy speech issue, comparing problem formulations and hypothesizing why separation system performance suffers when trained with mixtures of noisy speech. We demonstrate a gap in performance between systems trained in these two data paradigms by using simulated data---training models on the same mixtures but controlling whether the system has access to the clean or noisy oracle speech sources during training.

In addition, we propose a new objective function for more effectively training speech separation systems with synthetic mixtures of noisy speech signals. This function is designed to exploit the general orthogonality of unrelated audio signals along with the relative inseparability of noise mixtures to minimize the effect of separation errors resulting from a part of the signal deemed to be inseparable.

The core contribution of this work is to articulate an issue with training data for noisy speech separation systems and demonstrate the impact it can have on performance. Our proposed objective presents a promising avenue for exploring solutions to the problem.

\section{Separation Formulations and Challenges}
\label{sec:formulation}

The basic problem formulation of single-channel speech separation is to produce time-domain estimates $\hat{s}_k(t)$ of speech signals $s_k(t)$ of length $T$ with sample index $t$ for each speaker $k$ of $K$ total speakers in a mixture $x(t)$ with each of those speakers talking simultaneously:
\begin{align}
x(t) = \sum_{k=1}^K s_k(t)
\end{align}
In nearly all state-of-the-art deep learning-based speech separation systems, models are trained with a loss function that encourages each $\hat{s}_k$ to become equivalent to $s_k$, requiring knowledge of the ground-truth speech signals. As such, these models are trained using synthetic mixtures, i.e. $x(t)$ is not a real recording of multiple talkers speaking simultaneously, but rather a digital sum of each of $K$ real speech recordings $s_k(t)$.

\subsection{Noisy Separation Formulations}
\label{ssec:noisyforms}

In many real-world situations with multiple people speaking simultaneously, a recording will include more than just clean speech, such as noise or reverberation~\cite{bengio2005machine}. In particular, we consider a formulation in which a mixture includes a noise signal $n(t)$ as well:
\begin{align}
x(t) = \sum_{k=1}^K s^\text{clean}_k(t) + n(t) \label{eqn:clean_form}
\end{align}
In this formulation, conventionally the noise signal is considered to be undesirable, so the target for this task remains simply the $s^\text{clean}_k(t)$ signals. Once again, the requirement of ground truth training targets means the mixture must be a synthetic combination of clean speech along with a separate noise signal.

Unfortunately the need for separate clean speech and noise signals greatly reduces the amount of training data available and precludes the ability for noisy in-domain training. Creating artificial mixtures using already-noisy speech signals opens the door to more sources of data, but the problem formulation becomes different. Rather than having a set $s^\text{clean}_k(t)$ of speech signals along with a noise signal $n(t)$, we have a set of noisy speech signals $s^\text{noisy}_k(t) = s^\text{clean}_k(t) + n_k(t)$ each consisting of a combination of unique speech and noise, which are summed to create the mixture
\begin{align}
x(t) = \sum_{k=1}^K s^\text{noisy}_k(t) = \sum_{k=1}^K \big[ s^\text{clean}_k(t) + n_k(t) \big] \label{eqn:noisy_form}
\end{align}
where we never have access to the ``true'' speech signal $s^\text{clean}_k(t)$.

Without access to the clean ground truth $s^\text{clean}_k(t)$, the network is instead trained with the noisy speech signals $s^\text{noisy}_k(t)$ as target. Though arguably more desirable for the separation network to produce $s^\text{clean}_k(t)$ than $s^\text{noisy}_k(t)$, it is not an inherently incorrect separation solution as long as the speech signals themselves have been separated, particularly when taking into account the possibility to then feed the noisy separated output into a de-noising speech enhancement network, something shown to be successful with synthetic noisy mixtures \cite{whamr}. However, this paradigm nevertheless likely has issues stemming from inseparability of the noise mixtures.

\subsection{Challenges of Noisy Data Formulation}
\label{ssec:noisyprobs}

Speech separation and speech enhancement systems generally rely on the spectro-temporal properties of the signals. Speech and noise can be separated based on their differing statistical properties, and two speech signals can be separated based on their spectro-temporal structure, notably their sparseness~\cite{vincent2018textbook}. However, noise signals are not guaranteed to be spectrally sparse and can have similar statistical properties, particularly when datasets are collected in consistent environments. As a result, it may be very difficult to separate two noise signals. A noise separation task is introduced in~\cite{uss}, but its wide variety of noise sources and avoidance of ambient/environment tracks is not representative of background noise in conversational recordings from a consistent environment.

When training a network to separate a set of noisy sources $s^\text{noisy}_k(t)$, we are in essence asking the network to be able to discriminate each $s^\text{clean}_k(t)$ and $n_k(t)$ from $\{s^\text{clean}_l(t),\, n_l(t) \mid l \ne k\}$. But, discriminating the $n_k(t)$ from the $n_l(t)$'s may be disproportionately difficult compared to the other tasks, despite being unrelated to the separation of speech. As a result, it is likely that the performance of networks trained with $s^\text{noisy}_k(t)$ as target struggle disproportionately compared to networks trained with clean sources $s^\text{clean}_k(t)$ as target. In addition, even if the network can successfully separate the noise signals, it must correctly match them with the speech signal they originated from. As such, we believe that developing a training objective which minimizes the network's requirement to separate the noise will allow for training of more powerful models using datasets featuring synthetic mixtures of real noisy speech.

We do note that this formulation is very similar to the formulation in the MixIT work~\cite{mixofmix}, in which separation is learned using mixtures of mixtures. This work focuses on ground truth mixtures of speech rather than a combination of speech and noise. While this approach would address the noise-source assignment problem, we believe it would struggle to solve the overall noisy speech problem due to still requiring the noise signals to be separated. Our preliminary experiments using a similar approach were unsuccessful.

\section{Objective Function}
\label{sec:obj}

To solve this problem, we seek to develop an appropriate loss function that can encourage the network to produce estimates of clean source signals $\hat{s}_k(t)$ and a noise estimate $\hat{n}(t)$
\begin{align}
\hat{s}_k(t) & \approx s^\text{clean}_k(t) \label{eqn:est_srcs} \\
\hat{n}(t) & \approx \sum_{k=1}^K n_k(t) \label{eqn:est_noise}
\end{align}
given a set of ground truth noisy speech mixtures $s^\text{noisy}_k(t)$
\begin{align}
s^\text{noisy}_k(t) = s^\text{clean}_k(t) + n_k(t)
\end{align}
The estimates in equations \ref{eqn:est_srcs} and \ref{eqn:est_noise} are meant to estimate the components of the formulation presented in equation \ref{eqn:clean_form}, treating the sum of the individual noise signals $n_k(t)$ as one single noise source $n(t)$, despite the ground truth matching the formulation of equation \ref{eqn:noisy_form}.

\subsection{Design and Description}
\label{ssec:dnd}

Our approach to producing an appropriate loss function for this task is to modify the Scale-Invariant Signal-to-Distortion Ratio (SI-SDR) loss function~\cite{sisdr} to discount reconstruction errors that have been identified as belonging to the noise estimate. The goal is to partition the mixture implicitly---if a signal is known to be inseparable, it should be better to identify it as inseparable than get the separation wrong; if a signal is separable, it should be better to separate it than identify it as inseparable---relying on the assumption that speech is generally separable and noise is generally not.

One of the key assumptions we use for our loss function is that a sequence of audio samples can be modeled as a zero-mean stochastic process; so, while treated as vectors, two unrelated audio sources are approximately orthogonal. For this reason, from this point onward we will refer to the various components of a trial with vector notation: $\mathbf{s}^\text{clean}_k, \mathbf{s}^\text{noisy}_k, \hat{\mathbf{s}}_k, \mathbf{n}_k, \hat{\mathbf{n}}_k, \mathbf{x}\in \mathbb{R}^T$. As a result of this orthogonality property, as long as no speaker is repeated in a mixture and the noise recordings are not reused, the following is true:
\begin{align}
\langle \mathbf{a}, \mathbf{b} \rangle = 0\quad \forall\, \mathbf{a}, \mathbf{b} \in \mathcal{P}_\mathbf{x}, \  \mathbf{a} \ne \mathbf{b} \label{eqn:ortho}
\end{align}
where $\mathcal{P}_\mathbf{x} = \{\mathbf{s}^\text{clean}_1, \mathbf{n}_1,\dots,\mathbf{s}^\text{clean}_k, \mathbf{n}_k\}$ denotes the set of independent components present in the mixture $\mathbf{x}$. It should be noted that this formulation cannot consider reverberation to be a separate source from speech, as the independence assumption is not valid.

In addition, the above property leads to the following projection results for any $\mathbf{a}, \mathbf{b}, \mathbf{c} \in \mathcal{P}_\mathbf{x}$:
\begin{align}
\textit{proj}_{\mathbf{a}+\mathbf{b}}(\mathbf{a}+\mathbf{c}) & = \frac{\lVert \mathbf{a} \rVert^2}{\lVert \mathbf{a}+\mathbf{b} \rVert^2}(\mathbf{a}+\mathbf{b}) \label{eqn:projboth} \\
\textit{proj}_\mathbf{a}(\mathbf{a}+\mathbf{b}) & = \mathbf{a} \label{eqn:proj1}
\end{align}
These results can be useful for identifying the magnitude of shared components between two signals comprised of elements in $\mathcal{P}_\mathbf{x}$.

\begin{figure}[t]
\centering
\includegraphics[width=0.75\columnwidth]{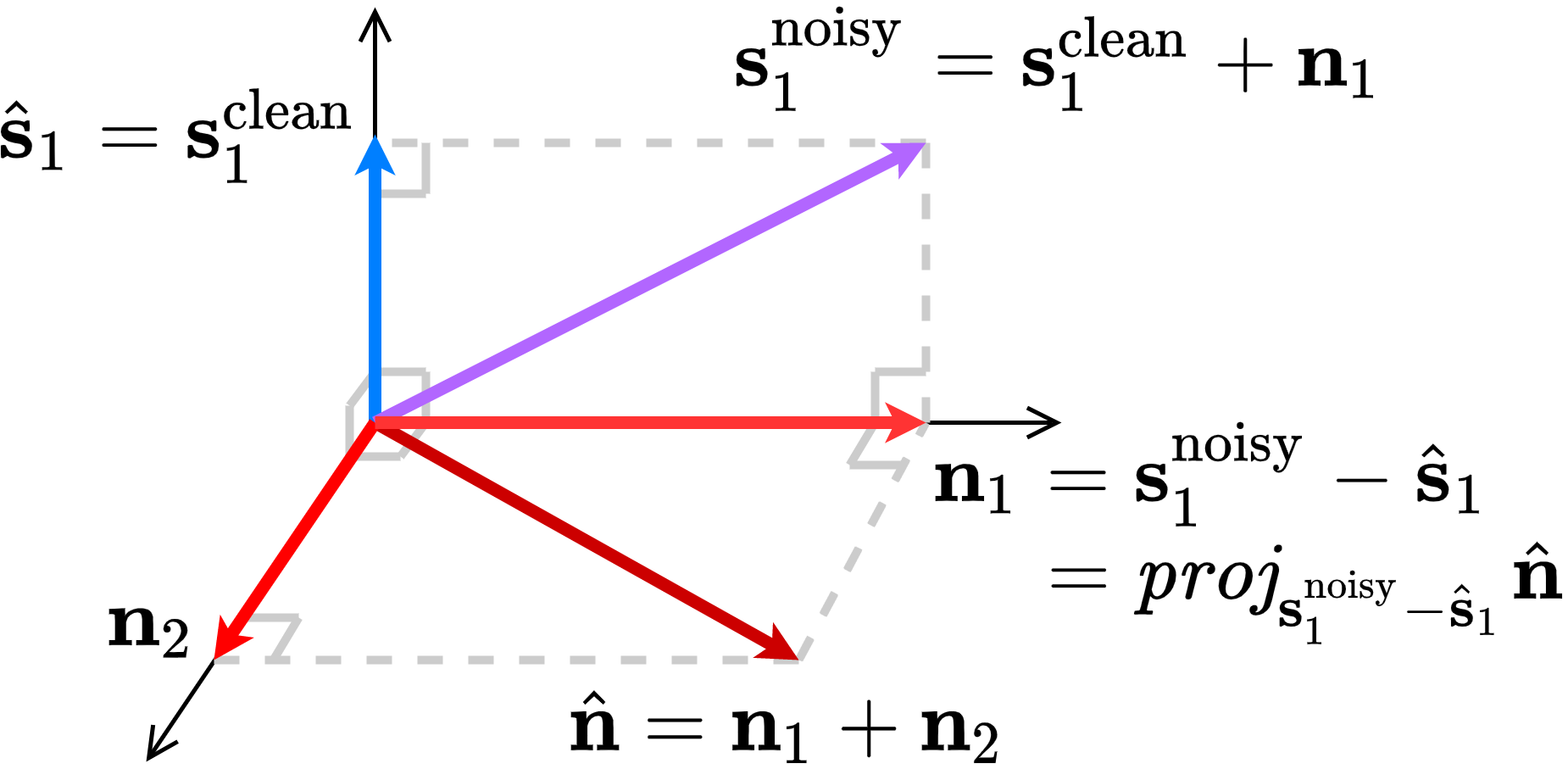}
\vspace{-8pt}
\caption{Illustration of the relationship between sources, noises, and their combinations.}
\label{fig:vecs}
\end{figure}

The basic SDR function of SI-SDR, while only having access to $\mathbf{s}^\text{noisy}_k$ as ground truth, would be defined as:
\begin{align}
\text{SDR}(\hat{\mathbf{s}}_k) \coloneqq 10 \log_{10}\frac{\lVert \mathbf{s}^\text{noisy}_k \rVert^2}{\lVert \mathbf{s}^\text{noisy}_k - \hat{\mathbf{s}}_k \rVert^2 } \label{eqn:sdr}
\end{align}
Our modified loss function, Estimated-Source-to-Separation-Error Ratio (ESSER) is defined as follows:
\begin{align}
\text{E} & \text{SSER}(\hat{\mathbf{s}}_k, \hat{\mathbf{n}}) \coloneqq \notag \\
& 10\log_{10}\frac{\lVert \hat{\mathbf{s}}_k \rVert^2}{\lVert (\mathbf{s}^\text{noisy}_k - \hat{\mathbf{s}}_k) - \lambda*\textit{proj}_{(\mathbf{s}^\text{noisy}_k - \hat{\mathbf{s}}_k)}\hat{\mathbf{n}} + \textit{proj}_{\hat{\mathbf{s}}_k} \hat{\mathbf{n}} \rVert^2} \label{eqn:esser}
\end{align}
ESSER differs from the SDR formulation in three main ways:
\begin{enumerate}[label=\roman*., itemsep=-1pt, topsep=4pt]
\item noise discount: $ \lambda*\textit{proj}_{(\mathbf{s}^\text{noisy}_k-\hat{\mathbf{s}}_k)}\hat{\mathbf{n}} $ \label{enum:discount}
\item orthogonality constraint:  $\textit{proj}_{\hat{\mathbf{s}}_k} \hat{\mathbf{n}}$ \label{enum:ortho}
\item separation encouragement: $\hat{\mathbf{s}}_k$ in numerator \label{enum:num}
\end{enumerate}
Modification (\ref{enum:discount}) exploits the projection in equation \ref{eqn:proj1} to discount separation errors that have been identified in the noise estimate, illustrated in Figure \ref{fig:vecs}, weighted by a tunable parameter $\lambda$ to prevent the network from over-categorizing the signal as noise. The orthogonality constraint modification (\ref{enum:ortho}) encourages the speech and noise estimates to not share content. Modification (\ref{enum:num}) replaces the oracle $\mathbf{s}^\text{noisy}_k$ with the estimated source $\hat{\mathbf{s}}_k$, to further encourage the network to separate as much of the waveform as possible.

\subsection{Issues}
\label{ssec:issues}


Two main issues arise that are essentially unavoidable in solutions to the noisy oracle source problem. Addressing them was necessary for ESSER-based systems.

\subsubsection{Scaling}
\label{sssec:scale}

Most state-of-the-art separation techniques output waveforms with semi-arbitrary scaling and must be re-scaled for proper objective function evaluation~\cite{tasnet,wavesplit,dprnn,shi2020sep,mixofmix}. The SI-SDR objective~\cite{sisdr} uses the result of equation \ref{eqn:proj1} to scale the signal such that the reconstruction error $ \mathbf{e}_k $ is orthogonal to the source $ \mathbf{s}^\text{clean}_k $. This does not work in the noisy source formulation, as the component $\mathbf{n}_k$ that is in $\mathbf{s}^\text{noisy}_k$ but not the estimate $\hat{\mathbf{s}}_k$ changes the projection to that of equation \ref{eqn:projboth}.

As a result, we scale the estimates by projecting the mixture onto them, which under-scales by the factor $ \lVert \hat{\mathbf{s}}_k \rVert^2 / \lVert \hat{\mathbf{s}}_k + \mathbf{e}_k \rVert^2 $. This is not optimal scaling, but converges to optimal as the reconstruction error decreases.

\subsubsection{Validation Tuning}
\label{sssec:tuning}

Another problem raised by this formulation is the issue of tuning performance. In this paradigm, we do not have access to data or metrics for the ``true'' task we are trying to complete. As such we cannot simply tune performance according to a held-out set.

In this case, we used SI-SDR on a noisy oracle validation set as a proxy function to tune $\lambda$. We swept $\lambda$ up from 0 by increments of 0.1, selecting the value prior to an SI-SDR decrease of greater than 0.667, signifying the network beginning to erroneously classify speech as noise.

\section{Experimental Configuration}
\label{sec:config}

\subsection{Datasets}
\label{ssec:data}

The data used for our experiments were new synthetic mixtures created using WHAM!\cite{wham}. While using synthetic data is not ideal, it is necessary for analyzing the variations in ground truth signals while controlling for other factors. We took the wsj0-2mix\cite{dpcl_wsj02mix} mixtures consisting of clean speech from the WSJ0 dataset~\cite{wsj0} and assigned each mixture two noise sources from the WHAM! noises, one for each source. The resulting samples can be configured for training or evaluation in a number of ways. First of all, the noises are scaled to be at a given signal-to-noise ratio (SNR) relative to their source, allowing simulations of the various SNRs present in real recordings. Secondly, the samples can be configured to mix the sources with their noise to produce two noisy samples, or they can be configured to produce two clean sources along with the mixture of noise, resulting in the ``noisy oracle'' and ``clean oracle'' formulations accordingly.

For our experiments, we used the 16 kHz sample rate and `min' configuration of the data. We evaluated datasets created with SNRs ranging from 25 dB to -5 dB in decrements of 5 dB, as well as clean speech and pure noise configurations.

\subsection{Evaluation}
\label{ssec:eval}

For evaluation we use the SI-SDR metric~\cite{sisdr}. For all speech signals across all conditions, we evaluate the raw SI-SDR value according to clean ground truth. This serves as a measure of the objective quality of the separated speech---as all dataset configurations have identical speech signals, they can be directly compared. For evaluating noise estimates, we use SI-SDR improvement.

\subsection{Model Configuration}
\label{ssec:model}

For all of our experiments we used a standard SI-SDR--trained TasNet-BLSTM~\cite{tasnet} with four Bi-directional Long Short-Term Memory (BLSTM) layers with 600 units in each direction. For the analysis and synthesis bases, we used 500 filters of length 5 ms with a shift of 2.5 ms.

Models were trained for 100 epochs using 4 second segments using the Adam~\cite{adam} algorithm with an initial learning rate of 0.001. The learning rate is decreased by a factor of two if the validation loss does not improve for three consecutive epochs. In addition, gradient clipping is performed with a maximum $\ell_2$ norm of 5. All networks were trained with either negative SI-SDR loss or negative ESSER loss in an utterance-level permutation-invariant manner~\cite{upit}.

We feel this setup is representative of state-of-the-art methods, as most modern architectures are based on a TasNet design~\cite{dprnn,mixofmix,shi2020sep}, with nearly all exceptions using an SDR training objective~\cite{wavesplit}.

\section{Results and Discussion}
\label{sec:results}

\begin{table}[tbp]
\centering
\caption{SI-SDR improvement [dB] of networks trained to separate only noise compared to only speech.}
\label{table:noisesep}
\vspace{2pt}
\begin{adjustbox}{max width=0.9\columnwidth}
\sisetup{table-format=2.1,round-mode=places,round-precision=1,table-number-alignment = center,detect-weight=true,detect-inline-weight=math}
\begingroup
\renewcommand*{\arraystretch}{0.9}
\begin{tabular}{cS[table-format=2.1,table-number-alignment = center]}
\toprule
Noise Separation SI-SDRi: & 0.395188 \\
Speech Separation SI-SDRi: & 15.3149 \\ \bottomrule
\end{tabular}
\endgroup
\end{adjustbox}
\vspace{-0.3cm}
\end{table}

The result of our experiment to separate pure noise is shown in Table \ref{table:noisesep}. An SI-SDR improvement of 0.4 is well below the standards of performance for speech separation, where an identical system reached 15.3 dB on clean speech. It is safe to say that this architecture is effectively unable to separate noise.

\begin{table}[tbp]
\centering
\caption{Noisy separation comparison across training objectives and ground truth with identical mixtures. The SI-SDR system trained with noisy ground truth sources serves as a performance floor, while the clean ground truth source SI-SDR system serves as a ceiling.}
\vspace{2pt}
\label{table:main}

\begin{minipage}[b]{0.63\linewidth}
\centering
\centerline{\robustify\bfseries
\renewcommand{\thefootnote}{\fnsymbol{footnote}}
\sisetup{table-format=2.1,round-mode=places,round-precision=1,table-number-alignment = center,detect-weight=true,detect-inline-weight=math}
\begingroup
\renewcommand*{\arraystretch}{0.7}
\begin{tabular}{S[table-format=2.1,table-number-alignment = center]|SS|S}
\toprule
\multicolumn{1}{c}{Datset} & \multicolumn{2}{c}{SI-SDR} & \multicolumn{1}{c}{ESSER} \\ \cmidrule(lr){1-1} \cmidrule(lr){2-3} \cmidrule(lr){4-4}
\multicolumn{1}{c}{SNR [dB]} & \multicolumn{1}{c}{Noisy} & \multicolumn{1}{c}{Clean} & \multicolumn{1}{c}{Noisy} \\ \midrule
$\infty$ & 15.3149 & 15.0018 & 15.3008 \\ 
15.0 & 11.8936 & 13.5454 & 11.8853 \footnotemark[2] \\
10.0 & 9.03732 & 12.0071 & 8.9923 \\
5.0 & 4.99705 & 10.3544 & \bfseries 5.71319 \\
0.0 & -0.149672 & 7.82664 & \bfseries 0.82951 \\
-5.0 & -9.00858 & 3.51661 & -9.28085 \footnotemark[2] \\ \bottomrule
\end{tabular}
\endgroup}
\vspace{2pt}
\centerline{(a) Separation SI-SDR [dB]}
\centerline{}
\end{minipage}
\hfill
\begin{minipage}[b]{0.35\linewidth}
\centering
\centerline{\robustify\bfseries
\renewcommand{\thefootnote}{\fnsymbol{footnote}}
\sisetup{table-format=2.1,round-mode=places,round-precision=1,table-number-alignment = center,detect-weight=true,detect-inline-weight=math}
\begingroup
\renewcommand*{\arraystretch}{0.7}
\begin{tabular}{S[table-format=1.1,table-number-alignment = center]S[table-format=2.1,table-number-alignment = center]}
\toprule
\multicolumn{2}{c}{ESSER} \\ \cmidrule(lr){1-2}
$\lambda$ & \multicolumn{1}{c}{Noise} \\ \midrule
0.0 & \multicolumn{1}{c}{--} \\ 
0.0 \footnotemark[2] & 6.32555 \\
0.0 & 6.0718 \\
0.3 & 2.1295 \\
0.3 & 0.675838 \\
0.2 \footnotemark[2] & -57.3634 \\ \bottomrule
\end{tabular}
\endgroup}
\vspace{2pt}
\centerline{(b) Noise Estimate}
\centerline{SI-SDRi [dB]}
\end{minipage}

\vspace{-0.3cm}
\end{table}
\renewcommand{\thefootnote}{\fnsymbol{footnote}}
\footnotetext[2]{Validation tuning of $\lambda$ failed for these systems and was selected \textit{ex post facto}. While these cases occurred outside the successful operating region of ESSER, it is worth noting the lack of robustness in the parameter tuning.}

\begin{figure}[htb]
%
\centering
\includegraphics[width=1.0\columnwidth]{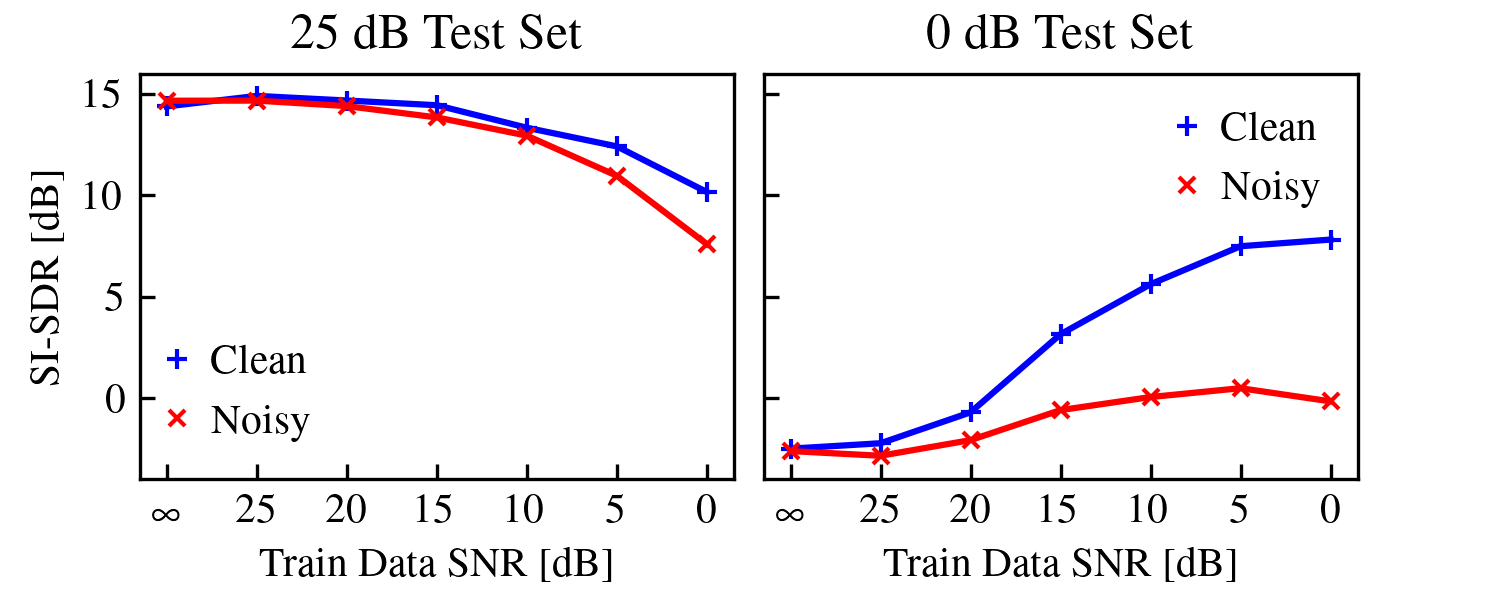}
\vspace{-20pt}
\caption{Evaluation of models trained with the SI-SDR objective with varying training data SNR across both ground truth paradigms.}
\label{fig:crosscond}
\end{figure}

Table \ref{table:main} shows the main results of our separation networks trained on noisy mixtures across the training data paradigms where the noise is or is not included in the oracle speech. As expected, the quality of speech produced by a model trained with clean ground truth decreases as more noise is added to the mixture. However, the performance of systems trained with noisy ground truth degraded more rapidly, with a growing gulf to the clean data paradigm as noise increased.

Figure \ref{fig:crosscond} shows two evaluation conditions, a nearly-clean 25 dB set and a noisy 0 dB set, evaluated with SI-SDR models trained over a variety of training sets with varying SNR in both ground truth paradigms. Interestingly, in the nearly-clean evaluation set, there was relatively minimal degradation as more noise was added to the training data, and little difference between the data paradigms. This suggests the noisy data paradigm has minimal impact on the network's ability to separate speech without noise.

However, in the noisy evaluation condition, the models trained on clean speech performed very poorly, with only the clean oracle source models significantly improving in performance as more noise was added to the training data. This suggests that adding noise to a model's training data does little to improve performance unless containing proper annotation as to what is speech and what is noise.

\begin{figure}[htb]
\centering
\includegraphics[width=1.0\columnwidth]{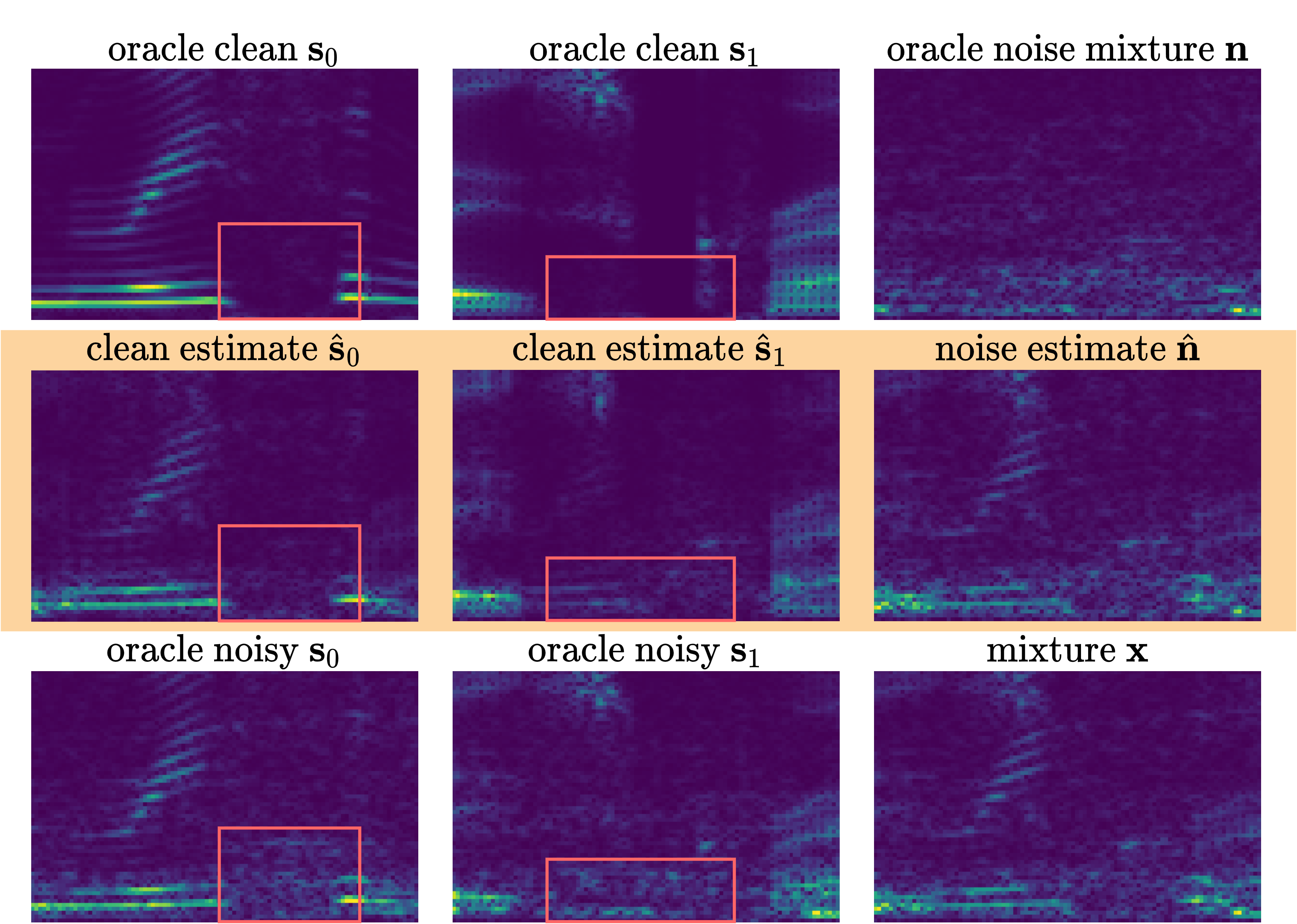}
\vspace{-20pt}
\caption{Sample section of magnitude spectra from the 0 dB evaluation set comparing ESSER system output to the oracle signals. The highlighted middle row is system output. The regions boxed in red are areas demonstrating noise suppression.}
\label{fig:specs}
\end{figure}

Table \ref{table:main} also includes the results of our experiments regarding training models with the negative ESSER loss function. These systems show modest gains over the SI-SDR noisy baseline in the 5.0 dB and 0.0 dB datasets, with sample output shown in Figure \ref{fig:specs}. The system fell apart at the highest level of noise, -5.0 dB, though the SI-SDR system did as well. At the lower levels of noise, the system performed best with $\lambda = 0$, effectively turning off the noise mitigation penalty, and accordingly matching the SI-SDR baseline.

In the ESSER systems that did not break down in very high noise, we see that the noise estimate does show an SI-SDR improvement, despite having no direct supervision, though the estimate itself resembles the mixture with only slight noise accentuation. Oddly enough, the systems where the noise discount penalty was removed ($\lambda = 0.0$), the noise estimate achieved approximately 6 dB improvement. An analysis of samples suggests that the orthogonality constraint (\ref{enum:ortho}) in conjunction with the scaling method and mask-based separation method led the network's noise estimate to produce components of the noise outside speech frequencies, not an overall estimate of the total noise.

\section{Conclusion}
\label{sec:conclusion}

We have demonstrated that there is a drop in performance in SI-SDR--trained separation systems trained on noisy data when the noise is present in the sources rather than being synthetically added to clean sources, impacting the effectiveness of in-domain training in noisy conditions. This degradation is likely due to requiring separation of two nearly-inseparable noise signals. As a result, we proposed an alternate loss function which exploits the inseparability of the noise to implicitly identify noise and minimize error contributed by failing to separate it. And, we have shown that this loss function can be used to train better systems on data with noisy sources which perform more effective separation on noisy mixtures.

\bibliographystyle{IEEEtran}
\bibliography{strings,refs}

\end{document}